\documentclass[twocolumn]{aastex701}
\usepackage{amsmath}
\usepackage{ulem}
\usepackage{enumitem,kantlipsum}

\newcommand{\be}{\begin{equation}}
\newcommand{\ee}{\end{equation}}
\newcommand{\ba}{\begin{eqnarray}}
\newcommand{\ea}{\end{eqnarray}}

\definecolor{blazeorange}{rgb}{1.0, 0.6, 0.2}
\definecolor{seagreen}{rgb}{0.18, 0.55, 0.34}
\definecolor{rufous}{rgb}{0.66, 0.11, 0.03}
\definecolor{royalfuchsia}{rgb}{0.79, 0.17, 0.57}
\definecolor{scarlet}{rgb}{1.0, 0.13, 0.0}
\definecolor{royalpurple}{rgb}{0.47, 0.32, 0.66}
\definecolor{darkblue}{rgb}{0, 0, 0.66}
\definecolor{violet}{rgb}{0.5,0.,0.5}

\begin{document}

\title{A milli-Tidal Disruption Event Model for GRB\;250702B: Main Sequence Star Disrupted by an IMBH}

\correspondingauthor{Jonathan Granot}
\author[0000-0001-8530-8941]{Jonathan Granot}
    \affiliation{Department of Natural Sciences, The Open University of Israel, P.O Box 808, Ra'anana 4353701, Israel}
    \affiliation{Astrophysics Research Center of the Open university (ARCO), The Open University of Israel, P.O Box 808, Ra’anana 4353701, Israel}
    \affiliation{Department of Physics, The George Washington University, Washington, DC 20052, USA}
    \email[show]{granot@openu.ac.il}

\author[0000-0002-5004-199X]{Hagai B. Perets}
    \affiliation{Department of Physics, Technion - Israel Institute of Technology, Haifa, 3200002, Israel}
    \affiliation{Astrophysics Research Center of the Open university (ARCO), The Open University of Israel, P.O Box 808, Ra’anana 4353701, Israel}
    \email{hperets@ph.technion.ac.il}

\author[0000-0003-0516-2968]{Ramandeep Gill}
    \affiliation{Instituto de Radioastronom\'ia y Astrof\'isica, Universidad Nacional Aut\'onoma de M\'exico, Antigua Carretera a P\'atzcuaro $\#$ 8701,  Ex-Hda. San Jos\'e de la Huerta, Morelia, Michoac\'an, C.P. 58089, M\'exico }
    \affiliation{Astrophysics Research Center of the Open university (ARCO), The Open University of Israel, P.O Box 808, Ra’anana 4353701, Israel}
    \email{rsgill.rg@gmail.com}

\author[0000-0001-7833-1043]{Paz Beniamini}
    \affiliation{Department of Natural Sciences, The Open University of Israel, P.O Box 808, Ra'anana 4353701, Israel}
    \affiliation{Astrophysics Research Center of the Open university (ARCO), The Open University of Israel, P.O Box 808, Ra’anana 4353701, Israel}
    \affiliation{Department of Physics, The George Washington University, Washington, DC 20052, USA}
    \email{pazb@openu.ac.il}

\author[0000-0002-9700-0036]{Brendan O'Connor}
    \altaffiliation{McWilliams Fellow}
    \affiliation{McWilliams Center for Cosmology and Astrophysics, Department of Physics, Carnegie Mellon University, Pittsburgh, PA 15213, USA}
    \email{boconno2@andrew.cmu.edu}  

\begin{abstract}
GRB\;250702B is the longest GRB recorded so far, with multiple gamma-ray emission episodes spread over a duration exceeding $25\;$ks and a weaker soft X-ray pre-peak $\sim1\;$day gradually rising emission. It is offset from its host galaxy center by $\sim5.7\;$kpc, and displays a long-lived afterglow emission in radio to X-ray.
Its true nature is unclear, with the two leading candidate classes of objects being a peculiar type of ultra-long GRB and a tidal disruption event (TDE) by an intermediate mass black hole (IMBH). Here, we consider the latter, mTDE origin. We model the afterglow data, finding a stratified external density profile $\propto r^{-k}$ with $k=1.60\pm0.17$,
consistent with Bondi accretion of the interstellar medium (of initial number density $n_{\rm ISM}=n_0\;{\rm cm^{-3}}$ and sound speed $c_s=c_{s,6}10^6\;{\rm cm\,s^{-1}}$) for which $n(r)\approx n_{\rm ISM}(r/R_{\rm B})^{-3/2}$ within the Bondi radius $R_{\rm B}$. Moreover, we use the implied density normalization to infer the IMBH mass within this model, finding $M_\bullet\approx\left(6.55^{+3.51}_{-2.29}\right)\times10^3\,n_0^{-2/3}\,c_{s,6}^{2}(1+\mathcal{M}^2)\;M_\odot$ where $\mathcal{M}\equiv v_{\rm BH}/c_s$ is the IMBH's Mach number relative to the ISM. 
Together with an upper limit on $M_\bullet<\frac{c^3}{G}\frac{t_{\rm MV}}{1+z} \lesssim5\times10^4\,M_\odot$ from the source-frame minimum variability time $t_{\rm MV,src}\!=\!\frac{t_{\rm MV}}{1+z}\!\approx\!0.5\;$s this implies $v_{\rm BH}\lesssim 28\,n_0^{1/3}\;{\rm km\;s^{-1}}$.
We show that a mTDE of a main-sequence star (but not of a white dwarf) can explain the duration and energetics of GRB\;250702B. The gradual rise to the peak may be caused by gradual circularization and accretion disk buildup, leading to an increase in the jet's power and Lorentz factor.

\end{abstract}


\section{Introduction} 

The unique event GRB\;250702B was detected by several different instruments onboard different satellites, including the \textit{Fermi}  
Gamma-ray Burst Monitor \citep[GBM;][]{Neights25GCN.40891,Neights25GCN.40931}, the \textit{Space Variable Objects Monitor} (\textit{SVOM}; \citealt{SVOM}), the \textit{Neil Gehrels Swift Observatory} \citep{Gehrels2004}, the \textit{Monitor of All-sky X-ray Image}  \citep[\textit{MAXI};][]{MAXI25GCN.40910}, Konus-Wind \citep{konusgcn} and the \textit{Einstein Probe} \citep[EP;][]{EP25GCN.40906}. 
Its main (or prompt) emission episode caused multiple Fermi-GBM triggers that were initially catalogued as separate GRBs (e.g., GRB\;250702B/D/E; \citealt{Neights2025}), but were later recognized to originate from the same sky location and are now collectively referred to as GRB\;250702B \citep{Levan2025}.

The prompt emission phase had a duration of $T_{90}\sim25$\,ks, the longest for any GRB, and it showed short timescale variability, with minimal variability times 
%in the range 
$t_{\rm MV}\sim1-10\;$s in different time intervals \citep{Neights2025}. It was also preceded by a soft X-ray ($0.5-4\;$keV) gradually rising pre-peak emission that was discovered by EP  \citep{EP25GCN.40906,EP250702a} and started about a day before the first trigger. The main emission episode was followed by a longer-lived emission in radio \citep{2025GCN.41059....1A,2025GCN.41147....1G,2025GCN.41145....1B,2025GCN.41061....1T,2025GCN.41053....1S,2025GCN.41054....1A}, infrared \citet{Carney2025} and X-ray \citep[][and references therein]{O'Connor2025}. The radio, infrared, and late ($>\!3\;$days) X-ray emission are consistent with an afterglow origin, from a forward external shock propagating into the external medium, which was powered by a relativistic 
jet launched by the source \citep[e.g.][]{Levan2025,O'Connor2025,Carney2025}. However, the early ($<\!3\;$days) X-ray emission shows significant short timescale variability (with $\Delta t/t$ as low as $\sim\!10^{-2.5}$) including sharp large-amplitude dips in the flux \citep{O'Connor2025}, which exclude an afterglow origin of the early-time ($<3\;$days) X-ray emission, which instead likely arise from internal dissipation in a relativistic collimated outflow driven by late time accretion onto the central source. 

The EP detection provided arcminute localization, which allowed the Swift-XRT detection that led to arcsecond localization \citep{Kennea25GCN.40919}. This in turn allowed near infrared observation and detection by the Very Large Telescope \citep[VLT;][]{Levan2025}, followed by \textit{Hubble Space Telescope} (\textit{HST}) observations that revealed an irregular host galaxy with the source offset from its center by $\sim0.7''$ \citep{Levan2025}. An observation with the \textit{James Webb Space Telescope} \citep{Gompertz2025} led to a host galaxy spectroscopic redshift of $z$\,$=$\,$1.036$, which implies a projected source offset from the galaxy center of $\sim5.7\;$kpc \citep{Carney2025} and an isotropic equivalent gamma-ray energy of $E_{\rm\gamma,iso}\gtrsim1.4\times10^{54}\;$erg \citep{Neights2025}. The soft X-ray pre-peak gradually rising EP emission had an isotropic equivalent energy of $E_{X,{\rm iso}}\!\,\!\gtrsim\!10^{52}\;{\rm erg}$
\citep{EP25GCN.40906,EP250702a}, and was likely beamed, although possibly less so than the main peak.

There is a very large dust extinction towards this source, both from our Galaxy 
($A_{\rm V}$\,$=$\,$0.847\;$mag; \citealt{Schlafly2011}) and 
from within its host galaxy \citep[$A_{{\rm V,}z}$\,$>$\,$5$ mag;][]{Levan2025,Carney2025,O'Connor2025,Gompertz2025}.
Together with the fairly high redshift, this would make it very difficult to detect any associated supernova, even if it was present and intrinsically luminous.

A scenario involving a stellar mass black hole is possible, and may naturally produce the very short minimal variability time, $t_{\rm MV}$ as low as $\sim\!1\;$s.
In particular many of the properties of GRB\;250702B are consistent with the long end of the ultra-long GRB population \citep[e.g.][]{Tikhomirova-Stern2005,Gendre2013,Levan2014,Greiner2015,Boer2015,Gendre2025}, whose origin is still unclear. 

Prolonging engine activity to $\sim\!10^4$–$10^5\,$s in collapsar-like models typically requires either finely tuned fallback envelopes or sustained accretion from a residual torus, and even then reproducing a smooth day-scale, gradually rising soft X-ray component preceding the main peak is challenging, since the jet must first drill through the stellar envelope while its power is still ramping up.

An alternative scenario featuring a tidal disruption event (TDE) by a stellar-mass black hole, dubbed micro-TDE or $\mu$TDE, was suggested as the origin of ultra-long GRBs \citep{Perets2016}, and is explored in detail as the origin of GRB\;250702B in \citet{BPG2025}.
White-dwarf - IMBH tidal disruption events have also been suggested as a possible origin for GRB\;250702B and other ultra-long GRBs \citep[e.g.][]{Eyles-Ferris2025,EP250702a}, however, as we also discuss below, this scenario 
%might be 
is significantly challenged if not excluded.

The significant offset from the host galaxy's center, along with the very short minimal variability time, strongly disfavors a regular TDE, caused by a super-massive black hole (SMBH). However, a TDE caused by an intermediate mass black hole (IMBH), i.e., a milli-TDE or mTDE, is still possible and is the scenario we study in this work. In particular, it is consistent with the $\sim5.7\;$kpc offset from the host galaxy center \citep{Carney2025}. Moreover, it may also accommodate the very short source frame minimal variability time, $\frac{t_{\rm MV}}{1+z}\sim0.5-5\;$s \citep{Neights2025}. 
A robust lower bound on significant variability of the high luminosity emission may be set by $r_g/c$, which in turn implies an upper limit of $M_\bullet<\frac{c^3}{G}\frac{t_{\rm MV}}{1+z} \lesssim5\times10^4\,M_\odot$ on the IMBH mass in a mTDE scenario \citep[see also][]{O'Connor2025}. 

In \S\,\ref{sec:afterglow} we perform a fit to the GRB\;250702B afterglow data, showing that the external density profile is consistent with a stratified external density profile expected for Bondi accretion. In \S\,\ref{sec:M_IMBH} we use the external density normalization derived in \S\,\ref{sec:afterglow} to infer the IMBH mass in a mTDE scenario (\S\,\ref{subsec:Bondi}) and consider different possible IMBH environments (\S\,\ref{subsec:BHL}). In \S\,\ref{sec:consistency} we discuss how a mTDE scenario can explain the timescales and energetics of the soft X-ray pre-peak emission (attributed to circularization, accretion disk buildup, and relativistic jet launching) and of the main emission (attributed to a fully-developed accretion-powered jet).
In \S\,\ref{sec:dis} we discuss our conclusions.

\section{Afterglow Modeling}
\label{sec:afterglow}

Here we fit the radio, infrared, and late ($>\!3\;$days) X-ray data \citep[for the data compilation, see][]{Levan2025,O'Connor2025,Carney2025} to a standard afterglow model from \citet{Gill-Granot-23}. This model assumes a uniform narrow jet of half-opening angle $\theta_j$, isotropic equivalent kinetic energy $E_{\rm k,iso}$, initial Lorentz factor $\Gamma_0\geq\theta_j^{-1}$, a power-law external mass density $n(r)=n_d(r/R_d)^{-k}$ with pivot radius $R_d=10^{18}\;$cm and density normalization $n_d=n_{0,d}\;{\rm cm^{-3}}$. It considers the joint dynamics of both the forward shock that propagates ahead of the ejecta into the external medium, as well as the reverse shock that propagates through the ejecta, to calculate the dynamical evolution of the contact discontinuity separating the two shocked regions \citep[for further details see][]{Gill-Granot-23}. Synchrotron radiation from the shocked material behind the two shocks is calculated using the standard afterglow theory that assumes a fraction, $\epsilon_B$, of the internal energy in magnetic fields and a fraction, $\epsilon_e$, in relativistic electrons with a power-law energy distribution with index $p$. The same quantities for the reverse shock are shown with a subscript `RS'. 

Observations were corrected for extinction ($A_{\rm V}$) and photo-electric absorption ($N_{\rm H}$) from within our galaxy, where the latter was also accounted for in the host galaxy \citep[see][]{O'Connor2025}. The afterglow fit separately accounts for extinction intrinsic to the host galaxy or local environment of the source.

\begin{figure}
    \centering
\includegraphics[width=1.0\columnwidth]{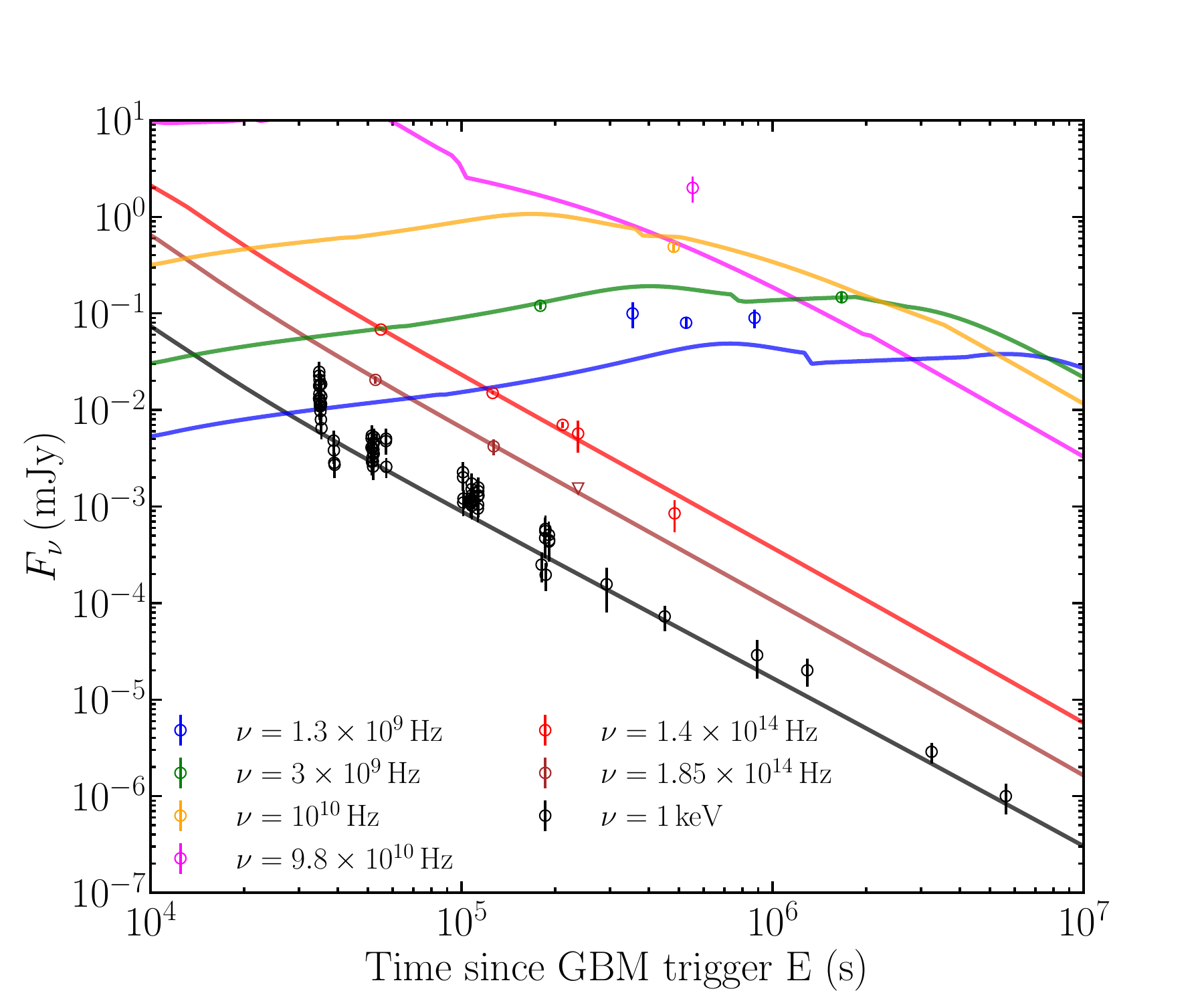}
    \caption{\label{fig:broadbandspec}
    Best-fit model lightcurve obtained from an MCMC fit to the multiwavelength afterglow data of GRB\,250702B. The model parameter posterior distributions are shown in Fig.\,\ref{fig:fit-posteriors}. The bump in the radio lightcurves at $10^5\,{\rm s}\lesssim t \lesssim 10^6\,{\rm s}$ is produced by a contribution from the reverse shock emission.
    }
\end{figure}

\begin{figure}
    \centering
    \includegraphics[width=1.0\columnwidth]{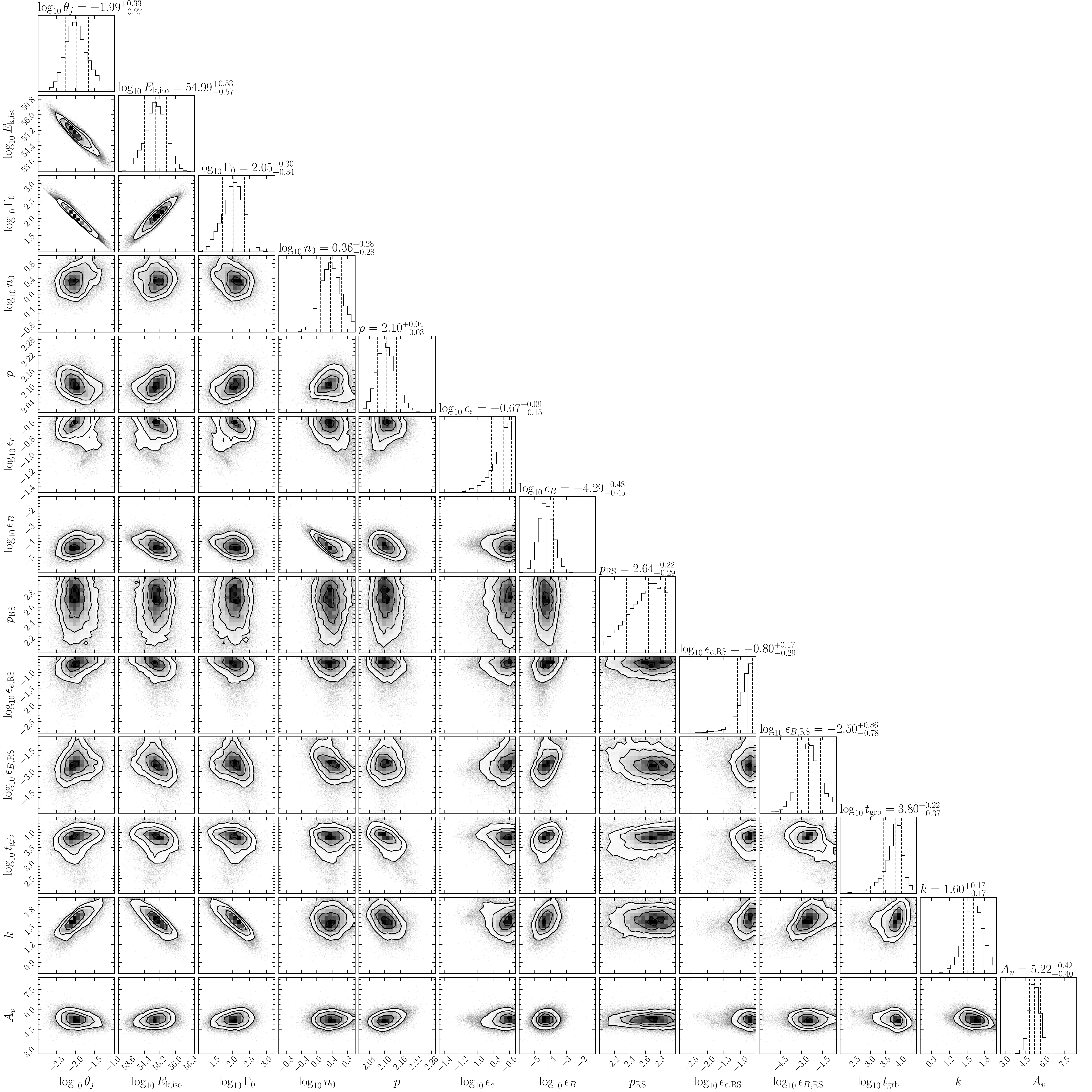}
    \caption{Model parameter posterior distributions obtained from an MCMC fit to the GRB\,250702B multiwavelength afterglow data using the afterglow model of \citet{Gill-Granot-23}. 
    }
    \label{fig:fit-posteriors}
\end{figure}

Figure\,\ref{fig:broadbandspec} shows the Markov Chain Monte Carlo (MCMC) fit to the multi-wavelength observations, where the lightcurves are obtained for the best-fit parameters as shown in the posterior distributions of parameters in Figure\,\ref{fig:fit-posteriors}. The X-ray and NIR light curves are completely explained by emission from the FS. The radio lightcurves, on the other hand, receive some contribution from the RS emission that produces the mild bump at times $t\sim10^5-10^6$\,s. Overall, the model lightcurves are consistent with observations.

The MCMC posteriors (Fig.\,\ref{fig:fit-posteriors}) show that, although there are the usual degeneracies between $E_{\rm k,iso}$, $n_d$ and the microphysical parameters, the external density slope $k$ is relatively well constrained and prefers a stratified profile ($k=1.60\pm0.17$) over a uniform medium ($k=0$). Values $0\!<\!k\!<\!2$ are also commonly inferred for long GRBs with well-monitored afterglows \citep[e.g.][]{Kouveliotou+13,Liang+13,Gompertz+18,Zhou+20}, so this result by itself does not rule out collapsar-like progenitors. However, in the context of an off-nuclear IMBH it is naturally interpreted as evidence for a Bondi/Bondi–Hoyle–Lyttleton–like inflow with $n(r)\propto r^{-3/2}$. The inferred beaming-corrected kinetic energy remains in the range $E_k\!\sim\!10^{50}$–$10^{51}\,$erg for reasonable choices of $\epsilon_e$ and $\epsilon_B$, which we later compare to the accretion-powered jet budget.

The best-fit value for the external medium density power-law index is $k=1.60\pm0.17$. 
This is consistent with the expectation of $k=3/2$ in the scenario of Bondi accretion of external medium onto an IMBH (see \S\,\ref{sec:M_IMBH}).

\section{Inferring the IMBH Mass} 
\label{sec:M_IMBH}

\subsection{A Simple Picture: an IMBH at rest in the ISM}
\label{subsec:Bondi}

Since the source is offset from the center of its host galaxy by $\sim5.7\;$kpc, and appears to lie in the galaxy's disk, it is likely surrounded by the interstellar medium (ISM), of number density $n_{\rm ISM}=n_0\;{\rm cm^{-3}}$, rest mass density $\rho_{\rm ISM}\approx n_{\rm ISM}m_p$ and sound speed $c_s=c_{s,6}10^6\;{\rm cm\,s^{-1}}$. Here we make the simple assumption that the IMBH is at rest relative to the ISM, while at \S\,\ref{subsec:BHL} we examine alternative scenarios.
Under our assumption here, an IMBH would accrete the surrounding ISM in a quasi-spherical Bondi-like flow. Such a flow is established within the Bondi radius,
\begin{equation}
\label{eq:RB}
R_{\rm B} = \frac{2GM_\bullet}{c_s^2} = 0.86\,M_{\bullet,4}\,c_{s,6}^{-2}\;\textrm{pc}\;,
\end{equation}
where $M_\bullet=M_{\bullet,4}10^4\;M_\odot$ in the IMBH mass.
Within $r<R_{\rm B}$ matter is accreted radially at close to the local Keplerian velocity, $v\approx v_{\rm K}=\sqrt{GM_\bullet/r}\propto r^{-1/2}$, 
such that a steady state constant $\dot{M}_{\rm acc}=4\pi r^2\rho(r)v(r)$, implies a density profile $\rho(r)\!\propto n(r)\!\propto r^{-3/2}$ (with $k=3/2$), or $n(r)\approx n_{\rm ISM}\max[1,(r/R_{\rm B})^{-3/2}]$. 

For a given ISM density $n_{\rm ISM}$, the density normalization within the stratified region ($r<R_{\rm B}$) is determined by the value of the Bondi radius $R_{\rm B}$, which in turn depends on the IMBH mass, $M_\bullet$. Since this density normalization is determined by our fit to the afterglow data, it can therefore be used to infer $M_\bullet$, which is of great interest. 
In \S\,\ref{sec:afterglow} we have derived $\log_{10}(n_{0,d})=0.36\pm0.28$, which when requiring $n_{\rm ISM}=n(R_{\rm B})$ implies a Bondi radius of $R_{\rm B}=R_d(n_d/n_{\rm ISM})^{2/3}=R_d(n_{0,d}/n_{0})^{2/3}\approx\left(0.56^{+0.31}_{-0.19}\right)\,n_0^{-2/3}\;$pc, and in turn an IMBH mass of
\begin{equation}\label{eq:M_IMBH}
M_\bullet\approx\left(6.55^{+3.51}_{-2.29}\right)\times10^3\,n_0^{-2/3}\,c_{s,6}^{2}\;M_\odot\;.
\end{equation}

This expression makes explicit that our mass estimate is most sensitive to the external density normalization and the sound speed, with $M_\bullet\propto n_0^{-2/3}c_s^{2}$. A denser or cooler ISM at fixed afterglow normalization implies a lower $M_\bullet$, whereas a more rarefied or hotter ISM pushes $M_\bullet$ upward. In this sense, the afterglow fit constrains a combination $n_0^{2/3}c_s^{-2}M_\bullet$, and Eq.~(\ref{eq:M_IMBH}) is best viewed as a one-parameter family of solutions parameterized by the (not directly measured) ISM properties.

\vspace{0.25cm}
\noindent\uline{\textbf{Consistency checks}}:
The value of $M_\bullet$ in Eq.~(\ref{eq:M_IMBH}) is consistent with the upper limit from $t_{\rm MV}$ (i.e. $M_\bullet\lesssim5\times10^4\,M_\odot$) and with the ability to tidally disrupt a white dwarf ($M_\bullet\lesssim10^5\,M_\odot$) or main sequence star ($M_\bullet\lesssim10^8\,M_\odot$). We now show it is self-consistent with:\\ (i) the assumptions of a Bondi-like accretion flow, and (ii) the afterglow shock propagating in the stratified region ($r\!<\!R_B$) during the afterglow observations used in \S\,\ref{sec:afterglow}.
We emphasize, however, that while this upper-limit argument shows that white-dwarf disruption is in principle possible for the inferred mass scale, the more detailed timing and energetic considerations presented in \S\,\ref{sec:consistency} disfavour a WD progenitor for GRB\;250702B.

\vspace{0.15cm}
\noindent\textbf{(i)}
The density profile $n(r)\approx n_{\rm ISM}\max[1,(r/R_{\rm B})^{-3/2}]$ implies an enclosed mass of accreting ISM within the Bondi radius,
\begin{equation}
M_B\equiv M(<\!R_{\rm B})=\frac{8\pi}{3}R_{\rm B}^3\rho_{\rm ISM} = 0.132\,n_0\,M_{\bullet,4}^3\,c_{s,6}^{-6}\;M_\odot\;.
\end{equation}
This mass rapidly increases with $M_\bullet$, until it equals $M_\bullet$ at a critical mass $M_{\rm cr}$ for which $M_B(M_{\rm cr})=M_\bullet$, and corresponding radius $R_{\rm cr}=R_B(M_{\rm cr})$, given by
\begin{eqnarray}\label{eq:M_cr}
M_{\rm cr} &=& \sqrt{\frac{3c_s^6}{64\pi G^3\rho_{\rm ISM}}} = 2.75\times10^6\,c_{s,6}^3n_0^{-1/2}\;M_\odot\;,
\\ \label{eq:R_cr}
R_{\rm cr} &=& \sqrt{\frac{3c_s^2}{16\pi G\rho_{\rm ISM}}} = 237\,c_{s,6}n_0^{-1/2}\;\textrm{pc}\;.
\end{eqnarray}
We note that up to factors of order unity this corresponds to the Jeans scale ($M_{\rm cr}\!\sim\! M_J$, $R_{\rm cr}\!\sim\!\lambda_J$). Beyond this scale (i.e., for $M>M_{\rm cr}\sim M_J$), the mass within $R_B$ is dominated by the ISM rather than by $M_\bullet$, and the ISM becomes unstable to gravitational collapse on scales smaller than $R_B$. Therefore, the assumptions leading to Bondi accretion hold only for $M_\bullet<M_{\rm cr}$. This is indeed consistent with the value of $M_\bullet$ derived in Eq.~(\ref{eq:M_IMBH}), and is less constraining than the upper limit from $t_{\rm MV}$. 

\begin{figure}
    \centering
\includegraphics[width=1.0\columnwidth]{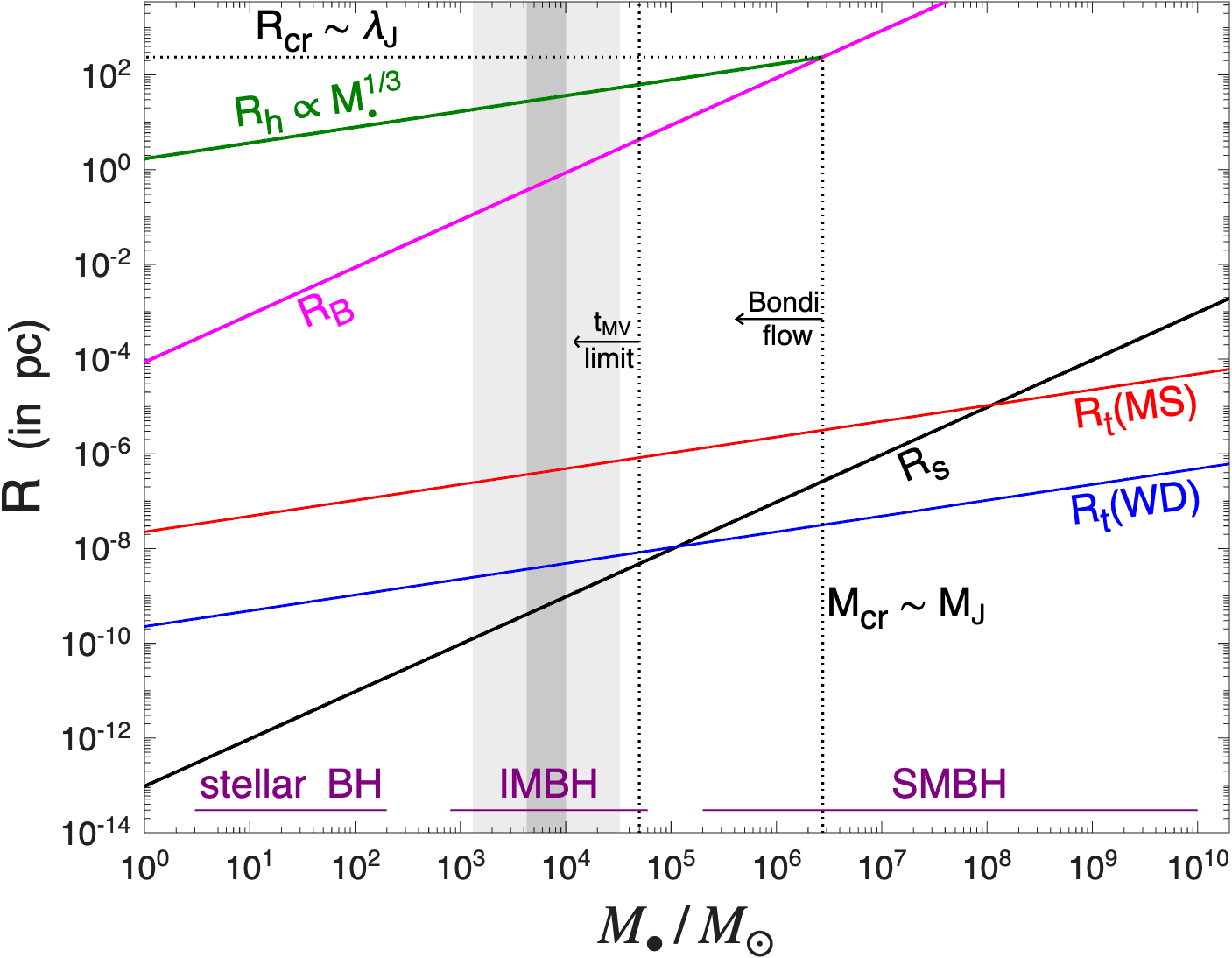}
\caption{Critical radii $R$ versus black hole mass $M_\bullet$: the tidal radius $R_t=R_*(M_\bullet/M_*)^{1/3}$ for a main-sequence (MS) solar-like star ($M_*=M_\odot$, $R_*=R_\odot$) and a white dwarf (WD; $M_*=M_\odot$, $R_*=0.01R_\odot$), the Schwarzschild radius
 $R_s = \frac{2GM_\bullet}{c^2}$,  the Bondi radius $R_B=\frac{2GM_\bullet}{c_s^2}$ (for $c_{s,6}=1$), and the radius of influence $R_h$ within which the ISM mass (for $n_0=c_{s,6}=1$) equals $M_\bullet$.
 Two upper limits on $M_\bullet$ are indicated by a vertical line and arrow. The narrow shaded gray region shows the inferred 1-$\sigma$ confidence interval on $M_\bullet$ (from Eq.~(\ref{eq:M_IMBH})) for $n_0^{-2/3}\,c_{s,6}^{2}=1$. The lighter and wider shading indicates a factor of 5 uncertainty in the value of $n_0^{-2/3}\,c_{s,6}^{2}$. Approximate mass ranges are indicated for stellar, intermediate mass (IMBH), and super-massive black holes (SMBH).
    }
    \label{fig:R-M}
\end{figure}

Figure~\ref{fig:R-M} shows relevant critical radii versus black hole mass. 
The narrow vertical shaded gray region shows the inferred 1-$\sigma$ confidence interval on $M_\bullet$ (from Eq.~(\ref{eq:M_IMBH})) for $n_0^{-2/3}\,c_{s,6}^{2}=1$ from our afterglow fit in \S\,\ref{sec:afterglow}. The lighter and wider shading indicates a factor of 5 uncertainty in the value of $n_0^{-2/3}\,c_{s,6}^{2}$. We note that other works \citep[e.g.][]{Carney2025,Gompertz2025} obtain a different external density normalization; while most works fix $k=2$, the afterglow data typically probes radii of the order of our pivot radius $R_d=10^{18}\;$cm, such that comparison to our $n_d=n(R_d)$ density normalization is meaningful. The inferred values vary by about an order of magnitude in either direction, which corresponds to a factor of $\sim5$ in our inferred value of $M_\bullet$, similar to the lighter, wider shading. 
Finally, a super-sonic motion of the IMBH relative to the ISM may lead to an increase in the estimated $M_\bullet$ (see \S\,\ref{subsec:BHL} and Eq.~(\ref{eq:M_IMBH_BHL}) below), but it is restricted by the $t_{\rm MV}$ limit, $M_\bullet\lesssim5\times10^4\;M_\odot$.

\vspace{0.15cm}
\noindent\textbf{(ii)} If the afterglow shock reaches $R_{\rm B}$ a transition will occur to a uniform medium, which would mimic a wind termination shock around a massive star GRB progenitor \citep[e.g.][]{Wijers01,Nakar-Granot07}, without a density jump by a factor of four in the latter case. For an afterglow flux decay $F_\nu\propto t^{-\alpha}$ this would lead to a flattening by $\Delta\alpha=0.5$ for $\nu_m<\nu<\nu_c$ but would hardly be noticed ($\Delta\alpha\approx0$) for $\nu>\max(\nu_m,\nu_c)$ \citep{GS02,Nakar-Granot07}. For an afterglow jet of initial isotropic equivalent energy $E_{\rm k,iso}$ and half-opening angle $\theta_j=10^{-1.5}\theta_{j,-1.5}$ this will occur before the jet break time $t_j$ if $E_{\rm k,iso}>M_Bc^2\theta_j^{-2}=2.36\times10^{56}\,\theta_{j,-1.5}^{-2}n_0\,M_{\bullet,4}^3\,c_{s,6}^{-6}\;$erg. Moreover, for a highly stratified external density profile, the jet break is very gradual \citep[e.g.][]{Granot07,DeColle+12}, and may easily be missed.  In our modeling the jet break occurs early, after a few hours, such that most of the afterglow observations are at $t>t_j$. In this case, when neglecting the jet's lateral spreading the $t_{\rm B}$ transition occurs at observed time $t_{\rm B}\approx78n_0\,E_{\rm k,iso,54.5}^{-1}\,M_{\bullet,4}^4\,c_{s,6}^{-8}\;$days. If $\nu_c$ was well above the Chandra energy range (which is at best only marginally valid), then the lack of a flattening in the X-ray lightcurve up to the second Chandra observation \citep{chandragcn2,Eyles-Ferris2025}, 
would imply $t_{\rm B}\!>\!65\;$days and in turn $M_{\bullet}\gtrsim9.5\times10^3E_{\rm k,iso,54.5}^{1/4}c_{s,6}^{2}\;M_\odot$.

For fast lateral spreading the Lorentz factor drops exponentially with radius beyond the jet break radius, $R_j$, where $R_j/R_B\approx0.056(E_{\rm k,iso,54.5}/n_0)^{2/3}\,M_{\bullet,4}^{-2}\,c_{s,6}^{4}$ and the jet becomes non-relativistic at a radius $\sim R_j(1-\ln\theta_j)$. In this case, the jet may become non-relativistic before reaching $R_B$ ($t_{\rm nr}<t_B$), but then the non-relativistic transition will be observed at
\begin{equation}
t_{\rm nr} \approx 100\,\frac{1-\ln\theta_j}{1-\ln 10^{-1.5}}E_{\rm k,51}^{2/3}n_0^{-2/3}M_{\bullet,4}^{-1}c_{s,6}^2\;\textrm{days}\;,
\end{equation}
where $E_{\rm k}=10^{51}E_{\rm k,51}\;$erg
is the jet's true kinetic energy. Note that the transition at $t_{\rm nr}$ should be apparent also above $\nu_c$, more clearly applying to the X-rays.
In this picture the  flux decay rate a $t>t_{\rm nr}$ \citep[but before reaching the deep-Newtonian regime, e.g.][]{Granot+06} would be $F_{\nu>\nu_{m,c}}\propto t^{-\frac{(15-4k)p-20+10k}{10-2k}}$ ($1.54\leq\alpha\leq2.11$ for $2.1\leq p\leq2.2$) or $F_{\nu_m<\nu<\nu_c}\propto t^{-\frac{(15-4k)p-21+8k}{10-2k}}$, which implies $F_{\nu>\nu_{m,c}}(t_{\rm nr}<t<t_B)\propto t^{-\frac{9p-5}{7}}$ ($1.99\leq\alpha\leq2.11$ for $2.1\leq p\leq2.2$) or $F_{\nu_m<\nu<\nu_c}(t_{\rm nr}<t<t_B)\propto t^{-\frac{9(p-1)}{7}}$ ($1.41\leq\alpha\leq1.54$ for $2.1\leq p\leq2.2$) and then $F_{\nu>\nu_{m,c}}(t_{\rm nr}<t_B<t)\propto t^{-\frac{3p-4}{2}}$ ($1.15\leq\alpha\leq1.3$ for $2.1\leq p\leq2.2$) or $F_{\nu_m<\nu<\nu_c}(t_{\rm nr}<t_B<t)\propto t^{-\frac{15p-21}{10}}$ ($1.05\leq\alpha\leq1.2$ for $2.1\leq p\leq2.2$). If $t_{\rm nr}\sim t_{B}$ then the flux decay would directly transition at this time to the latter slopes.

\subsection{Other Scenarios for an IMBH in the Host Galaxy}
\label{subsec:BHL}

Here we consider alternative scenarios for an IMBH within its host galaxy, such as: (i) having a significant speed $v_{\rm BH}\gtrsim c_s$ relative to the local ISM, (ii) being located at the center of a star cluster, or (iii) becoming episodically embedded in a gas-rich region that assembles a long-lived “mini-AGN” (mAGN) disc \citep{Rozner2025}.

\vspace{0.15cm}
\noindent\textbf{(i)} \uline{Fast moving IMBH}:
In scenario (i), the Bondi accretion flow is replaced by a Bondi-Hoyle-Lyttleton accretion flow. The simplest version of the latter is steady in the BH's frame and axisymmetric about the direction of $\textbf{\textit{v}}_{\rm BH}$. It is convenient to parameterize this flow using the BH's Mach number relative to the ISM, $\mathcal{M}\equiv v_{\rm BH}/c_s$. In the BH's frame, in the super-sonic case ($\mathcal{M}>1$), simulations show \citep[e.g.][]{Livio+1986,Blondin-Raymer2012,Xu-Stone2019} that a bow shock forms ahead of the central mass (BH), where the matter that remains bound accretes onto it in a flow that resembles a Bondi flow at $r<R_{\rm BHL}$ where\footnote{Note that $R_{\rm BHL}$ is of the order of the downstream Bondi radius $R_{B,d}$. For weak shocks ($\mathcal{M}\!\sim\!1$) both $R_{\rm BHL}\!\sim\!R_B\!\sim\!R_{B,d}$, where $R_{B,d}\sim R_B$ since the downstream and upstream sound speeds are comparable, $c_{s,d}\sim c_s$. For a strong shock it follows the scaling $R_B\propto c_s^{-2}\sim\frac{\rho}{p}$ where the pressure increases by a factor of 
$\sim\!\mathcal{M}^2$ while the density increases by a factor of $4$.}
\begin{equation}
R_{\rm BHL}\approx\frac{R_B}{1+\mathcal{M}^2} = \frac{2GM_\bullet}{c_s^2(1+\mathcal{M}^2)} = 0.86\,\frac{M_{\bullet,4}\,c_{s,6}^{-2}}{1+\mathcal{M}^2}\;\textrm{pc}\;,
\label{eq:R_BHL}
\end{equation} 
where $R_{\rm BHL}$ is the Bondi–Hoyle–Lyttleton radius, which generalizes the Bondi radius.
Up to factors of order unity, the density profile could be approximated by $n(r)\approx n_{\rm ISM}\max[1,(r/R_{\rm BHL})^{-3/2}]$, and our derived BH mass is generalized to
\begin{equation}\label{eq:M_IMBH_BHL}
M_\bullet\approx\left(6.55^{+3.51}_{-2.29}\right)\times10^3\,n_0^{-2/3}\,c_{s,6}^{2}(1+\mathcal{M}^2)\;M_\odot\;,
\end{equation}
i.e. it increases by a factor of $1\!+\!\mathcal{M}^2$ (or $\sim\!\mathcal{M}^2$ for $\mathcal{M}>1$) relative to pure Bondi accretion ($\mathcal{M}=0$).
Because $M_\bullet\propto n_0^{-2/3}c_s^{2}(1+\mathcal{M}^2)$ (Eq.~\ref{eq:M_IMBH_BHL}), higher upstream density and lower sound speed both act to \emph{reduce} the inferred $M_\bullet$ at fixed afterglow normalization, while larger IMBH–gas Mach number increases it through the $(1+\mathcal{M}^2)$ factor.

Using our upper limit on \(M_\bullet\) from the source-frame rapid variability time \(t_{\rm MV,src}\!=\!\frac{t_{\rm MV}}{1+z}\!\approx\!0.5\;\mathrm{s}\) of GRB\;250702B,
 $M_\bullet\lesssim5\times10^4\,M_\odot$, we obtain an upper limit on $\mathcal{M}$ and on $v_{\rm BH}$, 
\begin{equation}\label{eq:v_BH_UL}
\mathcal{M}\!\equiv\!\frac{v_{\rm BH}}{c_s} \lesssim 2.8\,n_0^{1/3}c_{s,6}^{-1}\ \ \Longrightarrow\ \ v_{\rm BH}\lesssim 28\,n_0^{1/3}\;{\rm km\;s^{-1}}\;.
\end{equation}

This upper limit is barely consistent with the escape velocity from the center of a globular cluster ($v_{\rm esc}\sim30-100\;{\rm km\;s^{-1}}$), where an IMBH may be naturally formed \citep{Gre+20}. Nonetheless, such an escaping IMBH with $v_{\rm BH}\sim v_{\rm esc}\lesssim30\;{\rm km\;s^{-1}}$ might still be consistent with the observations for GRB\;250702B. Even for an escape at a slightly higher velocity, the IMBH's velocity relative to the ISM $v_{\rm BH}$ may be damped over time because of dynamical friction with gas or stars.

\vspace{0.15cm}
\noindent\textbf{(ii)} \uline{IMBH at the center of a star cluster}:
If the IMBH resides in a star cluster, the ambient gas is set by the competition between stellar mass loss (e.g., AGB winds) and clearing by pulsar/UV feedback and/or ram pressure as the cluster moves through the host ISM. In old Milky Way globulars the steady-state gas content is generally very low, with typical central densities $n_{\rm gas}\!\sim\!10^{-3}$–$10^{-1}\ {\rm cm^{-3}}$ and stringent upper limits in several systems \citep{Fre+01,McDonaldZijlstra2015}; episodic retention (e.g. in core-collapsed/massive clusters) can temporarily raise $n_{\rm gas}$ to $\sim\!0.1$–$1\ {\rm cm^{-3}}$ before feedback re-clears the core \citep[e.g. see][and refs.\ therein]{Bobrick2025}. Higher, more sustained gas levels are plausible in young massive or nuclear clusters. The IMBH speed relative to the local gas is at most comparable to the cluster one-dimensional stellar velocity dispersion, $\sigma_\star\!\sim\!5$–$20~{\rm km\,s^{-1}}$ for old globulars and up to a few $10~{\rm km\,s^{-1}}$ in compact massive systems, implying $v_{\rm BH}\!\sim\!\sigma_\star$.
For warm gas ($T\!\sim\!10^4$\,K; $c_s\!\sim\!8$–$12~{\rm km\,s^{-1}}$), this gives $\mathcal{M}\!\equiv\!v_{\rm BH}/c_s\!\sim\!0.5$–$2$ (and larger $\mathcal{M}$ in colder gas). In this regime the flow is well described by Bondi–Hoyle–Lyttleton accretion, for which we have derived an upper limit on $v_{\rm BH}$ in Eq.~(\ref{eq:v_BH_UL}), $v_{\rm BH}\!\lesssim\!28\,n_0^{1/3}\,{\rm km\,s^{-1}}$, 
so a cluster-core IMBH with $\sigma_\star\!\sim\!10$–$25~{\rm km\,s^{-1}}$ is consistent with our constraints provided the gas is warm (or $n_0\!\gtrsim\!1$). Episodes of enhanced gas retention (or passages through denser clumps) would increase the normalization of $n(r)$ and, at fixed $k$, reduce the $M_\bullet$ required by the afterglow fit, while gas-poor phases do the opposite. For $n_{\rm gas}\!\sim\!0.1$–$10\ {\rm cm^{-3}}$ and $c_s\!\sim\!5$–$15~{\rm km\,s^{-1}}$, the Bondi radius (given in Eq.~(\ref{eq:RB})),
is $\sim 0.1$–$1$\,pc, large enough that the blast wave remains within the $n\!\propto\!r^{-3/2}$ zone over the epochs used in our fit, consistent with the measured $k=1.60\pm0.17$.

\vspace{0.15cm}
\noindent\textbf{(iii)} \uline{IMBH embedded in a gas-rich mini-AGN disc}:
An IMBH traversing or residing in a gas-rich clump can assemble a rotationally supported “mini-AGN” (mAGN) disc that sustains an elevated ambient density and mass-supply rate \citep{Rozner2025}. 
In the Bondi/BHL language, this raises the effective upstream density and can modify the sound speed. 
The relevant gravitational “capture radius”
that sets the inner $n\!\propto\!r^{-3/2}$ region is the Bondi–Hoyle–Lyttleton scale $R_{\rm BHL}$ (Eq.~(\ref{eq:R_BHL})), which reduces to the Bondi radius $R_{B}$ for $\mathcal{M}\!=\!0$. Inside $r\!\lesssim\!R_{\rm BHL}$ the steady inflow approaches $n(r)\!\propto\!r^{-3/2}$, so the afterglow normalization implies (cf. Eq.~(\ref{eq:M_IMBH_BHL})),
\begin{equation}\nonumber
M_\bullet\ \propto\ n_0^{-2/3}\,c_s^{2}\,\left(1+\mathcal{M}^2\right).
\end{equation}
\textit{Sign of the effect:} increasing the midplane density (larger $n_0$) or lowering the sound speed (smaller $c_s$) in an mAGN generally \emph{decreases} the $M_\bullet$ inferred from the same afterglow fit, whereas a larger relative shear (higher $\mathcal{M}$) \emph{increases} it through the $(1+\mathcal{M}^2)$ factor. Thus, a cool, dense mAGN midplane tends to lower the inferred $M_\bullet$ unless the IMBH–gas motion is sufficiently supersonic to compensate.

\section{A milli-TDE Main Sequence Star Model for GRB\;250702B}
\label{sec:consistency}

Here we discuss how an mTDE scenario can explain the different observational properties of GRB\;250702B or possibly of ultra-long GRBs (ULGRBs) in general. 
The classical GRB duration distribution is bimodal, consisting of short ($\sim10^{-2.5}-10^{0.5}\;$s) and long ($\sim10^{0}-10^{2.5}\;$s) GRBs, arising respectively from compact binary mergers and the collapse of massive stars \citep{Kouveliotou1993,Woosley1993,Eichler1989}. ULGRBs of durations $\gtrsim10^3\;$s \citep[e.g.][]{Gendre2013,Levan2014,Greiner2015,Boer2015,Gendre2025} form a small separate population, and appear to be distinct not only in the GRB duration distribution but also in their prompt emission spectral properties and host galaxy types.
(diverse prompt emission, spectral shapes, and host demographics) suggesting that ULGRBs may comprise multiple physical channels rather than a single progenitor class \citep{Gendre2013,Levan2014,Greiner2015}.

Their diverse prompt emission lightcurve shapes and spectra, along with mixed evidence on possible accompanying supernovae, may even suggest that ULGRBs may comprise multiple physical channels and/or subclasses rather than a single progenitor class \citep{Gendre2013,Levan2014,Greiner2015}.
GRB\;250702B extends this population to an extreme regime \citep{Neights2025} in both duration ($T_{90}\!\approx\!2.5\times10^4\,$s) and energetics ($E_{\gamma,\rm iso}\!\gtrsim\!1.4\times10^{54}\,$erg), while also exhibiting a day-scale, gradually rising soft X-ray emission episode (with $E_{\rm X,iso}\gtrsim10^{52}\;$erg; \citealt{EP250702a}). Any viable model must therefore explain not only the ultra-long duration, but also the structured temporal behavior across $\sim\,$five decades in time, from sub-second variability to $\sim\!10^5$\,s.

So far, we have shown that an mTDE model can, in principle, explain the observed short timescale variability of GRB\;250702B, as well as its external density profile and normalization, which provide a self-consistent estimate of $M_\bullet$. The main remaining observations that an mTDE model needs to explain are the duration ($>12\;$ks, where all quantities here are in the source's cosmological frame) and energetics of the main gamma-ray emission episode ($E_{\rm\gamma,iso}\gtrsim1.4\times10^{54}\;$erg), as well as the pre-peak gradually rising emission ($E_{\rm X,iso}\gtrsim10^{52}\;$erg) X-ray emission, which started about half a day earlier.

\subsection{Emission Timescales and Energetics}
A star of mass $M_*=M_{*,0}M_\odot$ and radius $R_*=R_{*,0}R_\odot$ is tidally disrupted near the tidal radius $r_t\equiv R_*(M_\bullet/M_*)^{1/3}$, or more precisely at $\eta^{2/3}r_t$ where $\eta\sim1$ for a main sequence star, and generally depends on the star's internal density profile \citep[e.g.,][]{Phinney1989,Lodato2009}. The depth of penetration relative to the tidal radius determines the disruption strength and is usually parameterized by the penetration factor, $\beta \equiv r_{\rm t}/r_p$, where $r_p$ is the peri-center distance. Deep encounters with $\beta\gtrsim1$ typically yield complete disruptions (in which about half of the star's mass becomes bound while the other half escapes, for an initial parabolic orbit), while shallower encounters ($\beta \lesssim 1$) can lead to partial disruptions and repeated stripping events \citep[e.g.,][]{Coughlin2019,Wang2021,Xin2024,Vynatheya2024}.

We consider a main-sequence (MS) sun-like star with $M_{*,0}\sim R_{*,0}\sim1$ and a white dwarf (WD) with $M_{*,0}\sim1$ and $R_{*,-2}=R_{*}/(0.01\,R_\odot)\sim1$. The time from the stellar disruption at $r_t$ to the first periastron passage at $r_p$ depends only on the properties of the disrupted star (and not on $M_\bullet$) and is approximately given by
\begin{equation}
\begin{split} 
    t_1 \;=\; \sqrt{\frac{\pi^2R_*^3}{2G M_*}}
  =  &\,\left\{ 
   \begin{array}{ll}
        3.54\!\times\!10^3 R_{*,0}^{3/2} \, M_{*,0}^{-1/2}
    \;{\rm s} & \hspace{0.05cm}\text{(MS)}\;,   \\
    \vspace{-0.25cm}\\
       3.54\, R_{*,-2}^{3/2} \, M_{*,0}^{-1/2}
    \;{\rm s} & \hspace{0.05cm}\text{(WD)}\;.
    \end{array} \right. 
\end{split}
\label{eq:tmin1}
\end{equation}

The most bound material acquires a semi-major axis $a_{\rm min}\approx \eta^{4/3}A_\beta^{2/3}r_t^2/2R_*$ and an orbital time
\begin{equation}
\begin{split}
     t_{\rm min}\,\approx\; 
     &\,\left\{ 
      \begin{array}{ll}
        3.5\times 10^4 \,A_{\beta,-1}\,
    \eta^2 \, M_{\bullet,4}^{1/2} \, R_{*,0}^{3/2} \, M_{*,0}^{-1}
    \;{\rm s} & \hspace{0.05cm}\text{(MS)}\;,   \\
    \vspace{-0.25cm}\\
       3.5\times 10^1 \,A_{\beta,-1}\,
    \eta^2 \, M_{\bullet,4}^{1/2} \, R_{*,0}^{3/2} \, M_{*,0}^{-1}
    \;{\rm s} & \hspace{0.05cm}\text{(WD)}\;.
    \end{array} \right. 
\end{split}
\end{equation}
where the parameter $A_\beta=10^{-1}A_{\beta,-1}$ \citep{BPG2025} depends on the star's response to the tidal field, ranging from 1 for the `frozen in' approximation to
$\sim\beta^{-3}$ if energy is efficiently dissipated near $r_p$, which we consider to be more realistic (e.g., \citealt{Guillochon2013,Stone2013,Coughlin-Nixon2015}). 
For a deep encounter with $\beta\sim3$ and $A_\beta\sim\beta^{-3}$ we have $t_{\rm min}\sim3.6\;$hr for a main sequence star, while a WD may not form an accretion disc at all and instead directly plunge in since $r_p\sim3.4r_g$ may be smaller than $r_{\rm ISCO}$. Since the most bound material falls back first, $t_{\rm min}$ sets the onset time of accretion. 

We can see that for a WD (even with a marginal $\beta\approx 1$), the characteristic timescales are far too short compared to GRB\;250702B \citep[see also][]{O'Connor2025}. A more detailed analysis, making explicit use of the observed intra-episode spacing $\Delta t_{\rm ep}\!\approx\!2.8\times10^{3}\,$s, the total prompt duration, and the energetics, shows that WD–IMBH encounters are in strong tension with the data even before invoking detonation limits. Interpreting $\Delta t_{\rm ep}$ as an orbital or fallback-based clock selects $t_{\rm min}\!\sim\!20$–$60\,$s, which is natural for low-mass WDs but then implies engine lifetimes that are much shorter than the observed multi-hour prompt phase. Conversely, choosing $t_{\rm min}$ large enough to power a hours-long engine is incompatible with the recurrence timescale. Together with the limited mass that can be stripped in non-detonating passages, this effectively rules out a WD–IMBH origin for GRB\;250702B, and from this point onward we therefore focus on a main-sequence progenitor in the mTDE framework, defined as MS-mTDE.

\vspace{0.25cm}
\noindent\uline{\textbf{GR precession, self-intersection, circularization}}:
At $r_p=r_t/\beta$ the general-relativistic (GR) apsidal precession per orbit is
\begin{equation}
\Delta\varpi =  \frac{6\pi r_g}{(1+e)r_p} \approx 3\pi\,\frac{r_g}{r_p}
%\;=\;  \approx 3\pi\,\beta\,\frac{G M_\bullet^{2/3} M_*^{1/3}}{c^2 R_*}
\approx
%0.54^\circ\,
0.93\!\times\!10^{-2}\beta\;
\frac{M_{\bullet,4}^{2/3}M_{*,0}^{1/3}}{R_{*,0}}\;,
\end{equation}
for a highly eccentric orbit ($e\approx1$). Efficient prompt circularization requires significant stream self-intersection (with a large relative velocity of the two streams at the intersection point, comparable to the local Keplerian velocity), which in turn requires precession exceeding the stream thickness, $\Delta\varpi \;\gtrsim H/R$, with debris-stream $H/R\!\sim\!10^{-2}$–$10^{-1}$. Because $\Delta\varpi$ is small for MS-mTDE encounters, prompt intersection is uncertain unless aided by pressure or Lense–Thirring precession. Nonetheless, each successive periastron pass leads to additional dissipation, which tends to decrease the semi-major axis $a$ and corresponding orbital time $P\propto a^{3/2}$, such that successive passages tend to take less time. If $P_i=q^{i-1}t_{\rm min}$ with some $q<1$ the the cumulative time for $m$ passages is $t_m = \sum_{i=1}^{m}P_i = \frac{1-q^{m-1}}{1-q}t_{\rm min}<\frac{t_{\rm min}}{1-q}$.
Hence, we adopt a conservative circularization multiplier
\begin{equation}
f_{\rm circ}\equiv\frac{t_{\rm circ}}{t_{\min}}\sim \mathcal{O}(1\text{--}10)\;,
\end{equation}
with lower values favored when precession or thickness aid intersection. Recent analytic and simulation work shows that stream self-intersection and stream–disk shocks can power early emission and that circularization may be delayed for weak precession typical of IMBH–MS encounters \citep{Hayasaki2016,Lu2020,Rossi2021,SteinbergStone2024}. For $\mu$TDEs by stellar-mass BHs, simulations likewise find that circularization efficiency depends sensitively on $\beta$ and flow thickness \citep{Kremer2023,Vynatheya2024}, and $f_{\rm circ}$ of order a few is plausible when vertical thickness or nodal precession aid intersections.

\vspace{0.25cm}
\noindent\uline{\textbf{The different timescales}}:
The circularization and viscous timescales of the debris disk also affect the effective accretion time. The local viscous time is $t_{\rm vis}\approx[\alpha h^2\Omega_{\rm K}(r)]^{-1}$ where $\Omega_{\rm K}=v_{\rm K}/r=(GM/r^3)^{1/2}$ is the local Keplerian angular velocity, $\alpha=10^{-1}\alpha_{-1}$ is the disk viscosity parameter and $h=H/R$ is the disk aspect ratio. 
If the infalling material circularizes near twice the peri-center distance (consistent with most of the dissipation occurring near $r_p$), then $R_{\rm circ} \simeq 2 r_p = 2r_{\rm t}/\beta$.
This leads to an accretion (i.e., viscous inflow) timescale of $t_{\rm acc}\approx t_{\rm vis}(R_{\rm circ})$, i.e.
\begin{equation}
t_{\rm acc}\approx\; 
4.5\!\times\!10^4 \,\beta^{-3/2}\,\alpha_{-1}^{-1}\,h^{-2}\,
    M_{*,0}^{-\frac{1}{2}} R_{*,0}^{3/2} \ {\rm s}\;.
\end{equation}

The large isotropic equivalent energy of the pre-peak X-ray emission, $E_{\rm X,iso}\gtrsim10^{52}\;$erg, suggests that it is beamed and arises from the relativistic jet. Its gradual rise towards the peak also supports a common origin with the main emission, which must arise much more clearly from a relativistic jet. It appears most likely that the pre-peak emission corresponds to the early stages of the circularization of the fallback accretion stream and the formation of an accretion disk. Therefore, its duration is expected to be of the order of $t_{\rm X,rise}\sim t_{\rm circ}+t_{\rm acc}$. As the accretion rate $\dot{M}_{\rm acc}$ gradually increases, it can support a larger magnetic flux near the IMBH, which can lead to a larger jet power.
Moreover, the initial circularization likely leads to a larger baryon loading in the jet and a lower Lorentz factor. As circularization is completed and the magnetic field anchored in the accretion disk near the BH reaches equipartition values it may inhibit significant baryon loading into the jet, allowing for a larger Lorentz factor during the main emission episode, and correspondingly stronger beaming of the emitted radiation (and a narrower jet with a smaller beaming factor $f_b\simeq\frac{1}{2}\theta_j^2$).

In such a scenario, we expect the duration of the main emission episode, $t_{\rm main}$, to correspond to the time over which $\dot{M}_{\rm acc}$ is near its peak value, roughly of the order of $t_{\rm min}+t_{\rm acc}=f_{\rm circ}^{-1}t_{\rm circ}+t_{\rm acc}$. 
Adopting here $A_\beta\approx\beta^{-3}$ and $\beta_{0.5}=\beta/10^{0.5}$ leads to
\begin{eqnarray} \nonumber
&t_{\rm X,rise}\sim t_{\rm circ}+t_{\rm acc}\;,\quad\ \ \,
t_{\rm main}\sim t_{\rm min}+t_{\rm acc}\;,\, 
\\ 
&t_{\rm circ}\approx 3.5\!\times\!10^4 f_{0.5}\beta_{0.5}^{-3}\,
    \eta^2 \, M_{\bullet,4}^{1/2} \, R_{*,0}^{3/2} \, M_{*,0}^{-1}
    \;{\rm s}\;,\, 
\\ \nonumber&t_{\rm min}\approx 1.1\!\times\!10^4\beta_{0.5}^{-3}\,
    \eta^2 \, M_{\bullet,4}^{1/2} \, R_{*,0}^{3/2} \, M_{*,0}^{-1}
    \;{\rm s}\;,\quad\ \ \,
\\ \nonumber
&\qquad t_{\rm acc}\approx\; 
3.2\!\times\!10^4 \,\beta_{0.5}^{-3/2}\,\alpha_{-1}^{-1}\,h_{-0.3}^{-2}\,
    M_{*,0}^{-\frac{1}{2}} R_{*,0}^{3/2} \ {\rm s}\;,\qquad\;
\end{eqnarray}
where $f_{0.5}=f_{\rm circ}/10^{0.5}$, $h_{-0.3}=h/10^{-0.3}\approx2h$. For a main sequence star the mass-radius relation follows $R_*\propto M_*^{s}$ with $s\approx0.8$ such that $t_{\rm circ}\propto R_{*}^{3/2}M_{*}^{-1}\propto M_*^{(3s-2)/2} \approx M_{*}^{0.2}$ only weakly depends on $M_*$, while $t_{\rm acc}\propto
    M_{*}^{-1/2}R_{*}^{3/2}\propto M_*^{(3s-1)/2} \approx M_{*}^{0.7}$ has a somewhat stronger dependence.

For a concrete fiducial example, taking a solar-type star ($M_{*,0}\!=\!R_{*,0}\!=\!1$), an IMBH mass $M_{\bullet,4}\!\sim\!0.7$ as suggested by Eq.~(\ref{eq:M_IMBH}), a moderately deep encounter $\beta\!\sim\!3$ ($\beta_{0.5}\!\approx\!0.95$), $\alpha_{-1}\!\sim\!1$ and $h_{-0.3}\!\sim\!1$, we obtain $t_{\rm min}\!\sim\!10^4\,$s, $t_{\rm acc}\!\sim\!{\rm few}\times10^4\,$s and $t_{\rm circ}\!\sim\!{\rm few}\times10^4\,$s for $f_{\rm circ}\!\sim\!{\rm few}$. These values are naturally of the same order as the observed source-frame durations of the pre-peak X-ray rise and main prompt phase, showing that the mTDE timescales for a main-sequence star around a $\sim\!10^4\,M_\odot$ IMBH are in the right ballpark without fine-tuning.

The fast rise to the main peak may be due to an instability in the accretion disk, which either boosted the jet power or increased its Lorentz factor and collimation, possibly by reducing the baryon loading as strong large-scale magnetic fields are established near the IMBH. 

The bright emission episodes (e.g., four Fermi-GBM triggers; \citealt{Neights2025}) within the main part of the emission may reflect cycles near the inner flow (e.g., through a magnetically arrested disk - MAD flux gating or interchange, or radiation-pressure thermal cycles), then a natural timescale is the local viscous (or thermal) time
$t_{\rm vis}=t_{\rm acc}(r/R_{\rm circ})^{3/2}$,
\begin{equation}
t_{\rm vis}\approx\frac{r^{3/2}}{\sqrt{GM_\bullet}\,\alpha h^2}=
1.24\!\times\!10^3\,r_{2}^{3/2}M_{\bullet,3.8}^{-1/2}\alpha_{-1}^{-1}\,h_{-0.3}^{-2}\;{\rm s}\;,
\end{equation}
where $r_{2}=r/(100\,r_g)$. Typical source frame emission episode durations of $\sim1-2\;$ks would correspond to radii $r\sim(85-140)r_g$.

Thus, if the bright episodes reflect quasi-cyclic behaviour in the inner flow (e.g. magnetically arrested disk flux accumulation and interchange, or radiation-pressure-driven thermal–viscous cycles), the observed $\sim\!10^3\,$s spacing can be interpreted as the local viscous/thermal time at $r\!\sim\!10^2\,r_g$. This interpretation is attractive because it ties the episode spacing directly to accretion physics rather than to an orbital clock at $r\!\gg\!r_t$, and it automatically scales with $M_\bullet$ in the same way as $t_{\rm acc}$.

\vspace{0.25cm}
\noindent\uline{\textbf{Main emission energetics}}:
As for the energetics of the main emission episode, our afterglow modeling requires a total true (beaming-corrected) kinetic energy in a narrow relativistic jet of $E_k\lesssim10^{51}\;$erg. For an accreted mass of $\sim0.5M_\odot$ this corresponds to a reasonable jet launching efficiency of $\eta_j\sim10^{-3}$.
An alternative parameterization following \citealt{BPG2025} requires an accreted mass of $M_{\rm acc}\approx 0.35\ \eta_{\gamma,-1}^{-1}\,\theta_{\rm j,-1.5}^2\,\eta_{\rm j,-2}^{-1}\ M_{\odot}$ where $\eta_\gamma=10^{-1}\eta_{\gamma,-1}=E_{\rm k,iso}/E_{\rm\gamma,iso}$, $\theta_j=10^{-1.5}\theta_{j,-1.5}$ is the jet half-opening angle and $\eta_j=10^{-2}\eta_{j,-2}=L_j/\dot{M}_{\rm acc}c^2$ is the jet launching efficiency. Here $\eta_j\sim10^{-2}$ and $\eta_\gamma\sim0.1$ are assumed. In the context of AGN, GRMHD simulations show that jet efficiencies span \(\eta_j\!\sim\!10^{-3}\)–\(10^{-1}\) depending on BH spin, magnetic flux, and whether the flow attains a magnetically arrested (MAD) state; values \(\gtrsim 10\%\) are obtained for high spin in MAD disks \citep{Tchekhovskoy2011,McKinney2012}. 
Such simulations assume a constant black hole spin, and are therefore not directly applicable to stellar-mass black holes where the accretion and jet launching can significantly change the BH spin if the accreted mass is comparable to $M_\bullet$, (e.g. \citealt{Wu2025} found that magnetically (Blandford-Znajek) powered jets in GRBs have a maximal efficiency of $\eta_{\rm j} \sim 0.015$). While the beaming factor of $f_b\approx\frac{1}{2}\theta_j^2\sim0.5\times10^{-3}$ used above ($\theta_{j,-1.5}$) is $\sim10$ times larger than the best fit value from our afterglow fit, it still compatible with it
(see also Eq.~(\ref{eq:fb_over_etaj_constraint}) below). 
Our best fit value of $f_b\sim10^{-4}$
leads to a similar required accreted mass for $\eta_\gamma\sim0.1$ and $\eta_j\sim10^{-3}$. 
Since the accreted mass in a TDE in much smaller than that of an IMBH, in our scenario the IMBH's dimensionless spin $a$ play an important role in determining the jet launching efficiency, which for relatively small values of $a$ is limited to $\eta_{\rm BZ}\lesssim0.25a^2$ \citep[e.g.][]{Tchekhovskoy2012}. This will in turn require a minimal spin parameter of $a\gtrsim0.06\eta_{j,-3}^{1/2}$.
Altogether, the energy requirements are reasonably met in a mTDE main-sequence star scenario.
Moreover, the beaming-corrected kinetic energy inferred from the afterglow modeling, $E_k\!\lesssim\!10^{51}\,$erg, is comfortably below the maximum budget available from accreting $\sim\!0.3$–$0.5\,M_\odot$ with $\eta_j\!\sim\!10^{-3}$–$10^{-2}$ and $\eta_\gamma\!\sim\!0.1$. In other words, the same parameter choices that reproduce the prompt $\gamma$-ray energetics are fully compatible with the independently inferred afterglow energetics, reinforcing the internal consistency of the mTDE interpretation.

\vspace{0.25cm}
\noindent\uline{\textbf{Explicit beaming–efficiency constraint}}:
One can reverse the argument in order to 
derive a limit on the beaming that is independent of the afterglow modeling.
The beaming-corrected prompt energy requires
\begin{equation}
M_{\rm acc}=\frac{f_b\,E_{\gamma,\mathrm{iso}}}{\eta_\gamma\,\eta_j\,c^2}.
\label{eq:Macc_req_MS}
\end{equation}
Requiring $M_{\rm acc}\le M_{\rm fb}=f_{\rm fb}M_*$, where $M_{\rm fb}$ and $f_{\rm fb}$ are the fallback mass and fraction, respectively, implies
\begin{equation}
f_b\le 6.4\times10^{-4}\eta_{j,-2}\eta_{\gamma,-1}
\!M_{*,0}\!\left(\frac{f_{\rm fb}}{0.5}\right)\!
\!
\left(\frac{1.4\!\times\!10^{54}\,\mathrm{erg}}{E_{\gamma,\mathrm{iso}}}\right),
\label{eq:fb_over_etaj_constraint}
\end{equation}
or correspondingly, $\theta_j\approx(2f_b)^{1/2}\lesssim0.036\;$rad.

\subsection{Expected Event Rates vs.\ Inferred Ultra-Long GRB Rates}
Ultra-long GRBs (ULGRBs; $T_{90}\!\gtrsim\!10^3$\,s) are rare in \textit{Swift}/BAT samples. A systematic search in the third BAT catalog finds $15$ ULGRBs, $\simeq2\%$ of BAT GRBs, with confirmed emission beyond $\sim\!10^3$\,s \citep{Lien2016}. Other compilations and case studies (e.g., \citealt{Levan2014,deWet2023}) similarly indicate a percent-level occurrence among prompt-detected events. There is also evidence that selection effects (in particular, the difficulty of detecting long, relatively faint emission tails) bias against ULGRBs, so this fraction should be regarded as a lower limit on the true underlying incidence \citep[e.g.][]{Levan2014}.

Using the local \emph{observed} long-GRB (LGRB) rate density $R_{\rm LGRB,obs}\!\sim\!1\!-\!2~{\rm Gpc^{-3}\,yr^{-1}}$ \citep[e.g.][]{Wanderman-Piran2010,Lan2019}, a naive observed ULGRB rate is
\begin{equation}
R_{\rm ULGRB,obs}\ \sim\ (2\!-\!4)\times10^{-2}\ {\rm Gpc^{-3}\,yr^{-1}}\;,
\end{equation}
simply scaling with the $\sim\!2\%$ ULGRB fraction. Correcting for beaming increases the intrinsic rate by $f_b^{-1}=2/\theta_j^2$. For a fiducial jet half-opening angle of $\theta_j\!\sim\!2^\circ$–$6^\circ$ (corresponding to $f_b\simeq\tfrac{1}{2}\theta_j^2\!\sim\!6\times10^{-4}$–$5\times10^{-3}$) the implied intrinsic rate is
\begin{equation}\label{eq:R_ULGRBint}
R_{\rm ULGRB,int}\ \sim\ 1\!-\!20\ {\rm Gpc^{-3}\,yr^{-1}}\;,
\end{equation}
where the numerical range primarily reflects the uncertainty in the typical ULGRB beaming angle, on top of the small-number statistics of the current sample and possible redshift evolution. This bracket should therefore be regarded as an order-of-magnitude estimate.

These rate estimates can now be compared with mTDE expectations. Dynamical models of star clusters and off-nuclear systems that retain an IMBH generally predict main-sequence
tidal-disruption rates per IMBH host of order
$\Gamma_{\rm mTDE}\sim 10^{-7}$--$10^{-5}\,{\rm yr^{-1}}$, with a
strong dependence on the cluster structure, mass, and IMBH occupation
fraction \citep[e.g.][]{StoneMetzger2016_TDEreview,Fragione2018_TDE_IMBH_GC}.
Converting to a volumetric rate requires the space density
$n_{\rm host}$ of suitable IMBH-bearing systems (massive star clusters, compact dwarfs, off-nuclear nuclei). Adopting a fiducial range\footnote{Corresponding
roughly to $\sim 10^2$--$10^3$ potential IMBH hosts per $L_\star$
galaxy, with a sub-unity occupation fraction.}
$n_{\rm host}\sim 10^6$--$10^8\,{\rm Gpc}^{-3}$ gives
\begin{equation}
R_{\rm mTDE} \simeq \Gamma_{\rm mTDE}\,n_{\rm host}
\sim 10^{-1}\!-\!10^{3}~{\rm Gpc}^{-3}~{\rm yr}^{-1},
\label{eq:RmTDE}
\end{equation}
which largely overlaps with the broad range of
$\sim\!10^{-2}-10^{2}\;{\rm Gpc^{-3}\;yr^{-1}}$  spanned by published calculations for different IMBH environments. We stress that $R_{\rm mTDE}$ is uncertain at the order-of-magnitude level, primarily because the census of IMBH-hosting systems is still poorly constrained.

Let $\xi\le 1$ denote the fraction of ULGRBs that are in fact IMBH mTDEs, and let $f_{\rm j}$ be the fraction of mTDEs that successfully launch relativistic jets, which would appear as ULGRBs if pointed close enough to our line of sight (the probability for which is $f_b$). Matching the intrinsic ULGRB rate to the mTDE pool
gives a simple bookkeeping relation,
\begin{equation}
f_{\rm j} \simeq \xi\,\frac{R_{\rm ULGRB,int}}{R_{\rm mTDE}}.
\label{eq:fj_bookkeeping}
\end{equation}
Here $R_{\rm ULGRB,int}=f_b^{-1}R_{\rm ULGRB,obs}$ already includes an assumed beaming correction
through the choice of $\theta_{\rm j}$; for a different fiducial opening angle both $R_{\rm ULGRB,int}$ and the inferred $f_{\rm j}$ would rescale proportionally.

Using $R_{\rm ULGRB,int}\sim 1$--$20~{\rm Gpc}^{-3}~{\rm yr}^{-1}$ and
$R_{\rm mTDE}\sim 0.1$--$10~{\rm Gpc}^{-3}~{\rm yr}^{-1}$ yields a broad allowed range,
\begin{equation}
10^{-1}\,\xi \lesssim f_{\rm j} \lesssim 1.
\end{equation}
If the true volumetric mTDE rate turns out to be at the high end of current theoretical expectations,
$R_{\rm mTDE}\sim 10^{2}-10^{3}\;{\rm Gpc}^{-3}~{\rm yr}^{-1}$, then proportionally smaller jetting fractions
$f_{\rm j}\sim 10^{-3}$--$10^{-2}$ with $\xi\sim 1$ would also be consistent. Thus, values $f_{\rm j}\sim 0.03$--$0.3$ are not strictly required; they are merely favored if one simultaneously adopts the upper end of $R_{\rm ULGRB,int}$ and the lower end of $R_{\rm mTDE}$ with
$\xi\simeq 1$.

In summary, within plausible (but currently broad) assumptions about beaming, jet production efficiency, and IMBH demographics, an mTDE of a main-sequence star by an IMBH can accommodate the observed rarity of
ULGRBs while also providing an intrinsic rate comparable to that inferred after beaming corrections. Improved constraints will come from (i) systematic ULGRB searches in survey data (e.g., BAT survey mode) \citep{Lien2016}, (ii) host-environment demographics of ULGRBs versus
TDEs, and (iii) direct IMBH census work in clusters and dwarf galaxies.

\section{Discussion and Summary}
\label{sec:dis}

\subsection{Summary of the main results}
We have explored the hypothesis that GRB\,250702B is powered by a milli-tidal disruption event (mTDE) in which a main-sequence (MS) star is disrupted by an intermediate-mass black hole (IMBH). Our multi-wavelength afterglow modeling favors a stratified external density profile with $k=1.60\pm0.17$, consistent with quasi-spherical Bondi/Bondi-Hoyle-Lyttleton (BHL) inflow where $n(r)\propto r^{-3/2}$ inside the capture radius. Using the afterglow-derived density normalization, we estimate the black-hole mass (see \S\,\ref{sec:M_IMBH} and Eq.~(\ref{eq:M_IMBH_BHL})),
\begin{equation}\nonumber
M_\bullet \approx \left(6.55^{+3.51}_{-2.29}\right)\times10^3\,
n_0^{-2/3}\,c_{s,6}^{2}\,(1+\mathcal{M}^2)\ M_\odot\;.
\end{equation}
This is comfortably below the independent upper bound $M_\bullet\!\lesssim\!5\times10^4\,M_\odot$ inferred from the source-frame minimum variability time $t_{\rm MV,src}\!\approx\!0.5$\,s, and consistent with disruption of an MS star outside the event horizon. Within this framework:

\vspace{0.1cm}\noindent
(i) The soft X-ray pre-peak rising emission likely arises from a relativistic jet (as suggested by its $E_{\rm X,iso}\!\gtrsim\!10^{52}\,$erg, which is too large for isotropic emission with plausible efficiencies),
during the circularization phase while an accretion disk builds up, leading to a gradual increase in the jet's power, Lorentz factor and degree of collimation. Its $\sim\,$half-day (source frame) timescale is naturally set by the sum of a circularization delay $t_{\rm circ}\!=\!f_{\rm circ}\,t_{\min}$ (with $f_{\rm circ}\!\sim\!{\rm few}$) and an accretion 
time $t_{\rm acc}\!\sim\!10^{4.5}$\,s at $R_{\rm circ}\sim2r_p$ for $\alpha\sim0.1$ and $h\sim0.3$\,--\,0.5;

\vspace{0.1cm}\noindent 
(ii) The multi-hour prompt $\gamma$-ray phase is naturally
attributed to a fully developed, narrowly collimated ultra-relativistic
jet, which emerges toward the end of the circularization stage when the
accretion rate $\dot{M}_{\rm acc}$ is close to its peak value; its
duration of order $t_{\min}+t_{\rm acc}$ matches the observed
few-hour engine activity.

\vspace{0.1cm}\noindent
(iii) Intra-episode spacings of $\sim\!10^3$\,s are consistent with local viscous/thermal times at radii $r\!\sim\!(50$-$150)\,r_g$. The afterglow fit requires a beaming-corrected kinetic energy $E_k\!\lesssim\!10^{51}$\,erg, which is achievable for jet efficiencies $\eta_j\!\sim\!10^{-3}$--$10^{-2}$ and accreted masses $M_{\rm acc}\!\sim\!{\cal O}(0.1$\,--\,$0.5)\,M_\odot$.

\subsection{A self-consistent physical picture}

A coherent picture emerges in which a $\sim\!M_\odot$ MS star is deeply
disrupted ($\beta\!\gtrsim\!{\rm few}$) by an IMBH of
$M_\bullet\!\sim\!10^4\,M_\odot$ embedded in warm ISM
($c_s\!\sim\!10~{\rm km\,s^{-1}}$) at an off-nuclear location. The
environment interior to the capture radius,
$R_{\rm B(HL)}\!\simeq\!2GM_\bullet/[c_s^2(1+\mathcal{M}^2)]$, naturally
provides $n(r)\!\propto\!r^{-3/2}$ as inferred from the afterglow
modeling. The pre-peak rising X-ray emission is produced during the
circularization process as an accretion disk starts to build up, and the
accretion rate $\dot{M}_{\rm acc}$ gradually increases along with the
jet’s degree of collimation and Lorentz factor. The main variable X-ray to gamma-ray emission arises once a powerful, narrowly collimated ultra-relativistic jet has formed and is powered by the near-peak
accretion rate.

The main, variable X-ray to gamma-ray emission occurs when the jet is fully developed, ultra-relativistic and narrowly collimated, near the peak $\dot{M}_{\rm acc}$. 
Late-time, smoother emission is afterglow-dominated (FS), with a modest radio bump from the RS.

\subsection{Alternative interpretations}
Three families of alternatives remain: 

\vspace{0.1cm}\noindent
\textbf{(i)} Ultra-long collapsars or other stellar-engine models: such can potentially yield long durations, but are hard pressed to produce
a $\sim$half-day gradually rising pre-peak soft X-ray emission, followed by a multi-hour peak, given that the jet needs to first bore its way through the progenitor massive star while its power is well below its peak value.

\vspace{0.1cm}\noindent
\textbf{(ii)} White-dwarf (WD) mTDEs: WD–IMBH tidal-disruption scenarios have been proposed for GRB\;250702B and other ultra-long GRBs \citep[e.g.][]{Eyles-Ferris2025,EP250702a} because WD structure can, in principle, provide short dynamical timescales and strong gravity. However, a simple analytic treatment anchored to the observed properties of GRB\;250702B shows that no self-consistent WD–IMBH configuration exists. Interpreting the intra-episode spacing
$\Delta t_{\rm ep}\!\approx\!2.8\times10^{3}\,$s as an orbital or fallback clock should be compared with typical values of $t_{\rm min}\!\sim\!20$--$60\,$s, which 
are orders of magnitude shorter than the observed multi-hour prompt duration even when allowing for generous circularization delays. Conversely,
choosing $t_{\rm min}$ large enough to power an hours-long engine would imply episode spacings far larger than observed. In addition, the cumulative mass that can be supplied by non-detonating partial-stripping passages over the observed time span is limited, so the beaming-corrected prompt energy can only be achieved for unrealistically narrow jets. Once basic detonation and ISCO constraints are also imposed, the allowed WD–IMBH parameter space for GRB\;250702B essentially vanishes.
We therefore regard a WD–IMBH origin as strongly disfavoured in comparison to the IMBH–MS mTDE scenario developed in this work, and refer the reader to \citet{BPG2025} for a complementary $\mu$TDE
interpretation tailored to GRB\;250702B.

\vspace{0.1cm}\noindent
\textbf{(iii)} micro-TDEs of a main-sequence star by a stellar mass black hole:
a detailed $\mu$TDE interpretation of GRB\,250702B is presented in our companion paper \citep{BPG2025}.
In brief, $\mu$TDEs (a stellar-mass BH/NS partially or fully disrupting a MS star) can naturally account for the sub-second variability and offer two viable routes to a day-scale pre-peak X-ray emission - the partial/repeating (dynamical) channel and the natal-kick (ballistic-delay) channel - both developed in \citep{BPG2025}. Here we note two points of tension that favor the IMBH mTDE picture for this event given the present afterglow dataset: \textbf{1}. the broadband afterglow fit prefers a stratified external density with $n\!\propto\!r^{-k}$ and $k\!\approx\!1.6$, suggestive of Bondi/BHL inflow onto an IMBH, whereas generic $\mu$TDE environments more often resemble uniform ISM or wind profiles unless special circumstances (ejecta/wind geometry) apply; and \textbf{2}. The combination of extreme isotropic energy and multi-hour prompt duration can push $\mu$TDEs toward relatively narrow jets and/or high jet efficiencies. Both caveats are model-dependent and explored in \cite{BPG2025}, which also lists observational discriminants (e.g., afterglow closure relations, very-late radio calorimetry, pre-peak spectroscopy). We therefore regard $\mu$TDEs as a competitive alternative and refer the reader to \citep{BPG2025} for a comprehensive treatment tailored to GRB\,250702B.

\subsection{Model limitations and caveats}
Our IMBH mass estimate is model-dependent: it assumes (a) that the blast wave is observed while still inside $R_{\rm B(HL)}$, (b) a quasi-steady Bondi/BHL profile, and (c) an external ISM characterized by $(n_0,c_s)$ that are not directly measured.  
We did verify the self-consistency of (a) and (b) in \S\,\ref{subsec:Bondi}, but this still doesn't guarantee their validity.
The circularization factor $f_{\rm circ}$ encapsulates complex GR hydrodynamics and may vary with $\beta$, BH spin, misalignment, and disk thickness; our choices ($f_{\rm circ}\!\sim\!{\rm few}$) are conservative but uncertain. Finally, the beaming factor and jet efficiency enter the prompt energy budget; both may evolve during the event.

\subsection{Observational tests and predictions}
\noindent
The mTDE picture makes several falsifiable predictions:

\vspace{-0.2cm}
\begin{enumerate}[leftmargin=0.4cm,labelwidth=0.4cm,itemsep=-0.15em]
\item \textit{Afterglow structure:} A transition from $k\!\approx\!3/2$ to $k\!\approx\!0$ is expected when the shock exits $R_{\rm B(HL)}$. This would produce a mild, achromatic light-curve flattening of $\Delta\alpha\!\approx\!0.5$ for $\nu_m\!<\!\nu\!<\!\nu_c$ (PLS G) and a similar $\Delta\alpha\!\approx\!0.5$ from $\alpha\approx0$ to $\alpha\approx0.5$ for $\nu<\mu_m<\nu_c$ (PLS D), but little change at $\nu\!>\!\max(\nu_m,\nu_c)$ (PLS H). Very late-time monitoring could reveal this transition.
\item \textit{Jet geometry:} A jet break may be very gradual in a stratified medium; deep late-time radio/X-ray observations may constrain $\theta_j$ via calorimetry, improving the beaming-corrected energy.
\item \textit{VLBI size and motion:} High-resolution VLBI observations in the radio can test for modest apparent expansion speeds and constrain the external density normalization independently of the SED fit.
\item \textit{Host environment:} Off-nuclear, dusty star-forming regions or star clusters at several kpc are natural locales for IMBHs; deep imaging/spectroscopy may reveal an associated compact stellar system.
\end{enumerate}

\subsection{Event rates and broader implications}
Accounting for selection biases, ULGRBs constitute a small fraction of GRBs, with an intrinsic rate plausibly $\sim$1-20~Gpc$^{-3}$\,yr$^{-1}$ after beaming corrections. Current estimates for IMBH TDEs in clusters and off-nuclear environments span $\sim$0.1-10~Gpc$^{-3}$\,yr$^{-1}$ with large uncertainties. If only a minority launch jets ($f_{\rm j}\!\sim\!\,10^{-1.5}$) and are beamed toward us, the observable rate becomes comparable to the detected ULGRB rate, suggesting that MS mTDEs could account for at least a subset of ULGRBs. Future wide-field, high-duty-cycle X-ray monitors paired with prompt radio follow-up will be key to building statistics.

\subsection{Outlook}
Several improvements can sharpen the mTDE interpretation:
(i) late-time radio calorimetry and VLBI size constraints; (ii) deeper X-ray monitoring to search for the predicted density-profile transition; (iii) polarimetry of the afterglow (jet magnetization); (iv) host-environment studies to identify a bound star cluster or dense association; and (v) detailed GRMHD simulations tailored to $M_\bullet\!\sim\!10^4\,M_\odot$ with realistic debris injection and BHL inflow. If confirmed, GRB\,250702B would establish IMBH mTDEs as a natural channel for ultra-long $\gamma$-ray transients and provide a new probe of wandering black holes and their environments at cosmological distances.

\medskip
\noindent\textit{In summary}, the MS mTDE scenario around an IMBH of $M_\bullet\!\sim\!10^4\,M_\odot$ self-consistently explains the half-day scale pre-peak X-ray emission, the multi-hour prompt X-ray to gamma-ray emission, the rapid variability, and the afterglow’s $n(r)\!\propto\!r^{-3/2}$ profile. While alternative models remain possible, the specific, testable predictions above offer a clear path to validating or refuting this interpretation with continued observations.

\vspace{0.25cm}\noindent
{\bf Acknowledgements}: 
We thank Eric Coughlin for useful discussions.
PB's work was funded by a grant (no. 2020747) from the United States-Israel Binational Science Foundation (BSF), Jerusalem, Israel and by a grant (no. 1649/23) from the Israel Science Foundation. BO acknowledges support from the McWilliams Fellowship in the McWilliams Center for Cosmology and Astrophysics at Carnegie Mellon University.

This work used resources on the Vera Cluster at the Pittsburgh Supercomputing Center (PSC). Vera is a dedicated cluster for the McWilliams Center for Cosmology and Astrophysics at Carnegie Mellon University. We thank the PSC staff for their support of the Vera Cluster.

The scientific results reported in this article are based on observations made by the Chandra X-ray Observatory. This research has made use of software provided by the Chandra X-ray Center (CXC) in the application package CIAO. This work made use of data supplied by the UK \textit{Swift} Science Data Centre at the University of Leicester. This research has made use of the XRT Data Analysis Software (XRTDAS) developed under the responsibility of the ASI Science Data Center (ASDC), Italy. This research has made use of data and/or software provided by the High Energy Astrophysics Science Archive Research Center (HEASARC), which is a service of the Astrophysics Science Division at NASA/GSFC.

\bibliography{bib250702}
%,references}
\bibliographystyle{aasjournal}
%\bibliographystyle{aasjournalv7}

%%%%%%%%%%%%%%%%%%%%%%%%%
\section{APPENDIX}\label{sec:appendix}
%%%%%%%%%%%%%%%%%%%%%%%%%

\end{document}